\documentclass{article}

\usepackage{amsfonts}
\usepackage{amssymb}
\usepackage{amsmath}
\usepackage{amsthm}
\usepackage{mathtools}
\usepackage{graphicx}
\usepackage{amscd}
\usepackage{mathrsfs}
\usepackage{tipa}

\begin{document}

\title{Is there any Trinity of Gravity,\\ to start with?}

\author{Alexey Golovnev\\
{\small \it Centre for Theoretical Physics, The British University in Egypt,}\\
{\small \it El Sherouk City, Cairo 11837, Egypt}\\
{\small  agolovnev@yandex.ru}} 
\date{}

\maketitle

\begin{abstract}
In recent years, it has been rather fashionable to talk about geometric trinity of gravity. The main idea is that one can formally present the gravity equations in different terms, those of either torsion or nonmetricity instead of curvature. It starts from a very erroneous claim that the Levi-Civita connection, and therefore the (pseudo-)Riemannian geometry itself, are nothing but an arbitrary choice. The point is that, as long as we admit the need of having a metric for describing gravity, the standard approach does not involve any additional independent geometric structures on top of that. At the same time, any other metric-affine model does go for genuinely new stuff. In particular, the celebrated teleparallel framework introduces a notion of yet another parallel transport which is flat. It gives us curious new ways of modifying gravity, even though very often quite problematic. However, in GR-equivalent models, we only get a new language for describing the same physics, in terms of absolutely unobservable and unpredictable geometrical inventions. For sure, one can always safely create novel constructions which do not influence the physical equations of motion, but in itself it does not make much sense and blatantly goes against the Occam's razor.
\end{abstract}

\section{Introduction}

The geometric trinity of gravity \cite{trinity} says that the gravitational interaction can be described in three different languages: curvature, torsion, or nonmetricity. In other words, they formally relate gravity to an arbitrary connection on the tangent bundle of our spacetime. At the same time, physically we deal with nothing but a metric manifold, and this metric geometry is what gives us the picture of the world around. Of course, any physical quantity is to some extent a matter of convention. However, the spacetime metric is fundamental for General Relativity in any incarnation one might imagine. We can either think of clocks and rods {\`a} la Einstein, or of geodesic lines followed by test particles, or even measurements of gravitational time dilation. However, the metric is always there. It is the basis of our modern understanding of gravity, both theoretically and observationally.

When the metric manifold is different from a simple Euclidean space (or Minkowski spacetime), the geometric relations, such as that of the circumference length to the radius, are modified, too. All this can be described by the Riemann tensor, that is the curvature tensor of the Levi-Civita connection. It is a property of the metric tensor only, with no need of introducing extra structures. The philosophy of trinity is based on an erroneous idea that the Levi-Civita connection was just an arbitrary choice. For sure, we are free to introduce new geometric structures, such as yet another connection or, equivalently, the torsion and/or the nonmetricity tensors. However, it is strange to do so without influencing the physical predictions.

In these notes, my aim is to elucidate the situation with alternative geometries for gravity. As long as we take the metric as physical, which would be hard to see otherwise, any new affine connection means enlarging the mathematical picture we use. In other words, we introduce some new entities. They might have some ideal dreams behind, such as quantising gravity, or defining conserved energy, or any other non-existent fantasies. However, if we restrict ourselves to GR-equivalent theories (with others almost always being problematic), then all dynamical predictions are the same, and therefore everything genuinely physical remains untouched, though with some unobservable structures introduced on top.

This paper is my contribution to the Proceedings of the 11th Mathematical Physics Meeting, September 2024 in Belgrade, Serbia. The plan is as follows. In the next Section 2 the standard Riemannian geometry is discussed, particularly stressing that the Levi-Civita connection is nothing of an arbitrary choice. Then I describe how I view the theory of General Relativity (GR) in Section 3. Section 4 is devoted to the basic ideas of teleparallel gravities, while Section 5 specifies them to GR-equivalent models. I share my views on the issues of conserved energy and quantum gravity in Section 6. Finally, in Section 7 I conclude.

\section{The standard (pseudo-)Riemannian geometry}

The very algebraic structure of Euclidean spaces (or Minkowski spacetimes for which I will often make no difference below) as vector spaces gives us the natural notion of two vectors being parallel at a distance. There exists a class of Cartesian coordinates in which the metric is represented by the standard matrix and every two parallel vectors do have their coordinate components equal. The corresponding parallel transport is unambiguous and corresponds to all the standard properties of ancient geometry. In particular, the parallelograms exist as closed figures.

At the same time, a manifold is a (Hausdorff and second-countable) topological space locally homeomorphic to a Euclidean space. In case of differentiable manifolds, we only allow for a class of coordinate charts which can be smoothly transformed one into another by going through the intersection of their images on the manifold. Given a metric $g_{\mu\nu}dx^{\mu}dx^{\nu}$ on a smooth manifold, its components can always be brought to the standard Euclidean (or Minkowskian $\eta_{\mu\nu}$ as below) form at a given point $x_*$, and up to the first order around it, 
$$g_{\mu\nu}(x) = \eta_{\mu\nu} +\sum_{\alpha,\beta} C_{\mu\nu\alpha\beta}\cdot (x-x_*)^{\alpha}(x-x_*)^{\beta} + {\mathcal O}((x-x_*)^3),$$
by a proper choice of coordinates (i.e. in the so-called normal coordinates). Intuitively, in a small enough region, a metric manifold is almost indistinguishable from a Euclidean space, including its natural notion of the parallel transport. The standard curvature tensor components at the chosen point $x_*$ are given by obvious linear combinations of the coefficients $C$, for what one can use the formulae (\ref{LC}, \ref{curve}) below. In other words, the Riemann tensor is a measure of impossibility to trivialise the metric beyond the linear order around a chosen point. It is not an independent thing.

To state it in yet another way, even though we fail to find a natural notion of being parallel at a distance, the natural procedure of parallelly transporting a vector or a tensor does come about from the approximate Euclidean structure. Since the vector space structure is only approximate, the results depend on the path taken, and therefore are not unique, unless performed in an infinitesimally small area. However, the Euclidean heritage does not change any scalar products, hence no nonmetricity, and moreover, it still allows for closed parallelograms, in infinitesimal scales, hence no torsion. It is then a textbook exercise, to see that it uniquely defines an affine connection, the Levi-Civita one,
\begin{equation}
\label{LC}
\mathop\Gamma\limits^{(0)}{\vphantom{\Gamma}}^{\alpha}_{\mu\nu}(g)=\frac12 g^{\alpha\beta}\left(\partial_{\mu}g_{\nu\beta}+\partial_{\nu}g_{\mu\beta}-\partial_{\beta}g_{\mu\nu}\right),
\end{equation}
with $g^{\alpha\beta}$ representing the dual metric, or the matrix inverse of $g_{\mu\nu}$. As always, the Einstein summation convention is adopted (see the end of \S 5 in the classic exposition of GR \cite{GR}).

\section{The Theory of General Relativity}

Let me repeat it again, the pseudo-Riemannian geometry of General Relativity is not just our choice. The real choice was in identifying gravity with the metric structure on a smooth manifold, and all the rest follows from it. Moreover, for the standard GR, the only dynamical variable is anyway a metric tensor, or in the usual understanding of physicists, the metric tensor components modulo diffeomorphisms. For the weird creatures like fermions, one could naturally add torsion to the game, {\`a} la Cartan, and it even enjoys some rather good motivation from supergravity and string theory, however there is nothing in the real-world data yet which would really require this torsion.

All in all, we describe gravity in terms of a non-trivial metric on the spacetime manifold. It automatically defines the Riemannian curvature tensor, irrespective of whether we want to admit that we work with curvature. Moreover, even the successful scheme for coupling matter to gravity goes in terms of introducing an arbitrary gravitational metric into the action functional for the matter fields. Then any test particle follows the geodesics, or the autoparallel curves of the Levi-Civita connection, with all the observable gravitational effects coming from the Riemann tensor. This is the only geometry we really need, and it is actually what is observable \cite{obsgeom}, at least in principle.

For really doing physics, we should restrict ourselves to only the things which can, at least as a matter of principle, be predicted and compared to the real world experience. From this point of view, the theory of General Relativity is nothing but the statement that
\begin{equation}
\label{theGR}
G_{\mu\nu}={\mathcal T}_{\mu\nu}
\end{equation}
with the left hand side being the Einstein tensor, and with the total energy-momentum tensor of all the matter contents in the right hand side. The equation (\ref{theGR}) can also be nicely presented as the one which follows from extremality of the action functional found by Hilbert,
\begin{equation}
\label{EH}
S=-\frac12 \int d^4 x \sqrt{-g} R\ + \mathrm{\ the\ action\ for\ matter\ contents},
\end{equation}
the only meaningful and nice action functional for reproducing the Einstein equations.

It is obvious that the Lagrangian density of $R=g^{\mu\nu}R^{\alpha}_{\hphantom{\alpha}\mu\alpha\nu}(\mathop\Gamma\limits^{(0)})$ does have second derivatives of the metric inside. However, for all the practical purposes of defining a classical theory, it just doesn't matter unless we are interested in some non-existent entities such as spacetime boundaries or gravitational energy. For deriving the equations of motion (\ref{theGR}), it is enough to assume stationarity of the action (\ref{EH}) with respect to variations of strictly finite support, so that even the overall global value of the action functional need not have any good meaning at all, i.e. the integral can safely be divergent. Of course, the bad problem would still appear if we suddenly wanted to do a path integral quantisation of gravity. The issues of quantum gravity I simply ignore for now. If anything, it is a different theory which should be defined separately.

For sure, one can say \cite{parcosm} that what I have described above is different from how Einstein viewed his theory. However, to a large extent, it was because he did not understand much about differential geometry, at least not at the time. A perfect window to how he was searching for the theory of gravity is his 1913 paper \cite{EinGross} with Grossman. It is composed of two parts, the physical part by Einstein, and the mathematical one by Grossman. The physical part starts from a very beautiful exposition of the equivalence principle. It was an ingenious observation of Einstein and, I would say, probably the only undoubtedly beautiful part of the paper. Given that locally our world's gravity is indistinguishable from simply being in a non-inertial frame, it is absolutely natural to seek a description in terms of free motion on a manifold.

Geometries can be different. However, we anyway need to measure distances and times, and therefore the minimal option would be to go for Riemannian geometry. Without proper geometric background in his mind, Einstein reached the correct conclusion about the need for the metric tensor by formally thinking of how one would write an action principle for particle motions. Then the physical part of the paper \cite{EinGross} continued into a dead end by taking analogies with electrostatics and looking for an energy of gravity, an illusion which Einstein would never be able to get rid of.

The mathematical part by Grossman \cite{EinGross} did basically have all what was needed, all the way to the Ricci tensor. It is even strange to see that it did not come to his mind to try a combination of $R_{\mu\nu}$ (in our modern notation) with $Rg_{\mu\nu}$. Maybe, the local conservation laws for matter fields did not ring a bell due to worries about gravitational contributions. All in all, not being able to reproduce the Newtonian limit, both authors concluded that most probably the idea of general covariance must be abandoned. In other words, the proper geometry was far out of sight.

It was difficult for Einstein to stop thinking about coordinates as something objectively constructed from rods and clocks, the view which had perfectly served his needs when doing special relativity and, alas, was not so good any longer. And when he had finally come to such understanding, he basically concluded that spacetime was nothing physical at all (see \S 3 of \cite{GR}). In this sense, he indeed did not think of gravity as geometry. This fact had even made it difficult for him to properly understand the nature of singularities (whether objective or merely coordinate ones) as can, for example, be seen from his reactions to the de Sitter solution \cite{EinDeS}. Moreover, irrespective of that, he was absolutely against the non-trivial empty spacetime \cite{EinDeS}, for it went against what he had called the Mach's principle.

In his views towards Mach and de Sitter, Einstein insisted \cite{EinDeS} that the physical qualities of space must be fully determined by matter, hence no independent $g_{\mu\nu}$ field should exist. Therefore, deep in the Einstein's mind, his theory of gravity was indeed quite different \cite{parcosm} from the beautiful geometric picture we have now. For sure, one can call the proper theory a "geometric dogma" \cite{enercan}, but is this change of attitude really what we need? Let's see the benefits and the price of teleparallel frameworks.

\section{The basics of teleparallel models}

In a metric-affine approach to gravity, one introduces yet another connection on the tangent bundle of the spacetime manifold. Recall that, at the same time, the Levi-Civita connection (\ref{LC}) is always there, due to the mere fact that we have the metric structure, for otherwise we do not even have any proper idea of how to couple the matter fields to gravity. In other words, in any metric-affine approach, we do introduce new variables to the theory. It can be done either in terms of the new connection $\Gamma^{\alpha}_{\mu\nu}$ (and the corresponding new notion of covariant derivatives $\bigtriangledown$) directly or via the tensors of torsion
\begin{equation}
\label{tor}
T^{\alpha}_{\hphantom{\alpha}\mu\nu} = \Gamma^{\alpha}_{\mu\nu}- \Gamma^{\alpha}_{\nu\mu}
\end{equation}
and/or nonmetricity
\begin{equation}
\label{nonmet}
 Q_{\alpha\mu\nu}=\bigtriangledown_{\alpha}g_{\mu\nu} 
= \partial_{\alpha}g_{\mu\nu} - \Gamma^{\beta}_{\alpha\mu}g_{\beta\nu} -\Gamma^{\beta}_{\alpha\nu}g_{\mu\beta}
\end{equation}
due to the general relation
\begin{equation}
\label{general}
\Gamma^{\alpha}_{\mu\nu}=\mathop\Gamma\limits^{(0)}{\vphantom{\Gamma}}^{\alpha}_{\mu\nu}+\frac12\left(T^{\alpha}_{\hphantom{\alpha}\mu\nu}+T^{\hphantom{\nu}\alpha}_{\nu\hphantom{\alpha}\mu}+T^{\hphantom{\mu}\alpha}_{\mu\hphantom{\alpha}\nu}\right)-\frac12\left(Q^{\hphantom{\mu\nu}\alpha}_{\mu\nu}+Q^{\hphantom{\nu\mu}\alpha}_{\nu\mu}-Q^{\alpha}_{\hphantom{\alpha}\mu\nu}\right)
\end{equation}
which is just as easy to derive as the classical case (\ref{LC}) itself.

At the same time, the curvature tensor
\begin{equation}
\label{curve}
R^{\alpha}_{\hphantom{\alpha}\beta\mu\nu}=\partial_{\mu}\Gamma^{\alpha}_{\nu\alpha}- \partial_{\nu}\Gamma^{\alpha}_{\mu\alpha} 
+\Gamma^{\alpha}_{\mu\rho}\Gamma^{\rho}_{\nu\beta} - \Gamma^{\alpha}_{\nu\rho}\Gamma^{\rho}_{\mu\beta}
\end{equation}
measures the non-integrability of the connection. Namely, it gives a change of a vector upon a parallel transport over a little closed loop. The teleparallel condition of $R^{\alpha}_{\hphantom{\alpha}\beta\mu\nu}=0$ means that the corresponding parallel transport is uniquely defined, at least modulo global topological issues. We consider a metric manifold with its Levi-Civita connection $\mathop\Gamma\limits^{(0)}{\vphantom{\Gamma}}^{\alpha}_{\mu\nu}$ and a teleparallel (i.e. flat, that is of vanishing curvature tensor) connection $\Gamma^{\alpha}_{\mu\nu}$ on top of it.

Any teleparallel model imposes a topological condition of global parallelisability on the spacetime manifold. Let's for simplicity assume even more, namely that the spacetime is topologically trivial. Over this topological ${\mathbb R}^4$, and given a flat connection, one can introduce a field of covariantly constant basis vectors $e_a^{\mu}$ by freely choosing it at one point and parallelly transporting to every other point of the manifold. Obviously, its dual basis $e^a_{\mu}$ is composed of covariantly constant 1-froms. The condition of $\partial_{\mu}e^a_{\nu}-\Gamma^{\alpha}_{\mu\nu}e^a_{\alpha}=0$ immediately shows that a teleparallel connection must have its coefficients of the Weitzenb{\"o}ck form:
\begin{equation}
\label{telcon}
\Gamma^{\alpha}_{\mu\nu}=e^{\alpha}_a \partial_{\mu}e^a_{\nu}.
\end{equation}

Modulo global general linear transformations of $e_a^{\mu}$ (reflecting the freedom of choosing the initial basis), the fields of covariantly constant bases and the flat connections are in one-to-one correspondence to each other. This is the geometrical meaning of the Weitzenb{\"o}ck connection \cite{meW, meRub}: there exists a basis of covariantly constant vectors. In the language of trinity, people usually introduce two special cases of teleparallel models, depending on which one of the tensors (\ref{tor}, \ref{nonmet}) is not vanishing. 

The classical one can be called the {\bf metric teleparallel} framework, for vanishing nonmetricity is assumed: 
$$R^{\alpha}_{\hphantom{\alpha}\beta\mu\nu}=0\qquad \mathrm{and}\qquad Q_{\alpha\mu\nu}=0.$$ 
In other words, the teleparallel connection can then be characterised in terms of the torsion tensor. With a severe abuse of common sense, people like to say that gravity is described in terms of torsion only. In this case, $Q_{\alpha\mu\nu}=0$, the matrix of $g_{\mu\nu}e_a^{\mu}e_b^{\nu}$ must take a constant value in the set of non-degenerate matrices, and it can be chosen to be $\eta_{ab}$, the standard Minkowski metric. So, for the metric teleparallel models, the nice and commonly used choice is to take the basis $e^{\mu}_a$ orthonormal. The simplest option is then to consider it as the dynamical variable while the metric is defined as
\begin{equation}
\label{metdef}
g_{\mu\nu}=\eta_{ab}e^a_{\mu}e^b_{\nu},
\end{equation}
and therefore it is also in the model, of course. This is the essence of the so-called "pure-tetrad" approach to the metric teleparallel gravity, even though one can also perfectly use non-orthonormal tetrads \cite{meW}, in particular the null ones \cite{nullt1, nullt2}. 

Another option is the {\bf symmetric teleparallel} framework. It means that
$$R^{\alpha}_{\hphantom{\alpha}\beta\mu\nu}=0\qquad \mathrm{and}\qquad T^{\alpha}_{\hphantom{\alpha}\mu\nu}=0.$$
According to the definition of the torsion tensor (\ref{tor}), it requires symmetry of the Weitzenb{\"o}ck connection (\ref{telcon}) in the lower indices. Hence, in a topologically trivial region of spacetime, the fundamental tetrad can always be represented in a gradient form:
\begin{equation}
\label{symtet}
e^a_{\mu}=\frac{\partial\zeta^a}{\partial x^{\mu}}.
\end{equation}
In other words, a generally covariant representation of symmetric teleparallel gravity would be in terms of the metric tensor $g_{\mu\nu}$ and four scalar fields $\zeta^a$ representing the Cartesian coordinates of the flat and torsion-free parallel transport structure \cite{mesym}.

At the same time, in all versions of teleparallel gravities, the basis $e^a_{\mu}$ in the formula (\ref{telcon}) corresponds to a particular flat connection and cannot be taken as arbitrary (unless the flat parallel transport is treated as an arbitrary thing in itself). In case one wants to use an arbitrary tetrad $h^a_{\mu}$, a "fully invariant" formulation \cite{full} is needed, and it can often be convenient for practical calculations, too \cite{mecov}. Then we must find the connection coefficients $\omega^a_{\hphantom{a}\mu b}$ in the new basis. It can obiously be done via
\begin{equation}
\label{tetpost}
\partial_{\mu}h^a_{\nu}-\Gamma^{\alpha}_{\mu\nu}h^a_{\alpha}+\omega^a_{\hphantom{a}\mu b} h^b_{\nu}=0,
\end{equation}
so that the preferred basis $e^a_{\mu}$ corresponds to $\omega^a_{\hphantom{a}\mu b}=0$. Unfortunately, the relation (\ref{tetpost}) is often called a "tetrad postulate", and the $\omega^a_{\hphantom{a}\mu b}$ components a "spin connection", especially when the tetrad $h^a_{\mu}$ is also orthonormal, even though this is nothing but the very same connection $\Gamma$ rewritten in the new basis $h_a^{\mu}$.

\section{The Teleparallel Equivalents of GR}

As has been mentioned above, each and every teleparallel representation of a gravity theory possessing the dynamical metric tensor does need an extra structure on top of the observable metric geometry. Actually, Einstein perfectly understood this fact, and his teleparallel ideas belonged to an (unsuccessful) attempt at unifying gravity and electromagnetism \cite{EinTor}. At the same time, the Teleparallel Equivalents of GR are given by action functionals which are fine-tuned in such a way as to make the choice of teleparallel structure absolutely arbitrary, hence the attributive noun "equivalent" for not producing anything new. Saying it in yet another way, the choice of the basis $e^a_{\mu}$ appears to be pure gauge, possibly restricted by the conditions (\ref{metdef}) or (\ref{symtet}) in the metric or symmetric versions respectively.

The fine-tuning is very easy to perform. Indeed, substituting the relation (\ref{general}) into the curvature tensor definition (\ref{curve}), we see that the vanishing curvature tensor of the teleparallel connection and the usual Riemann tensor are related to each other via a quadratic expression in terms of torsion and nonmetricity tensors. For simplicity, let me specify it to the two teleparallel cases of the trinity \cite{trinity}. One can easily find that
$$R(\mathop\Gamma\limits^{(0)})+2 \mathop\bigtriangledown\limits^{(0)}{\vphantom{\bigtriangledown}}_{\mu}T^{\mu}+\mathbb T=0 \quad \mathrm{where} \quad {\mathbb T} = \frac14 T_{\alpha\beta\mu}T^{\alpha\beta\mu}+\frac12 T_{\alpha\beta\mu}T^{\beta\alpha\mu}-T_{\mu}T^{\mu} \quad \mathrm{and} \quad T_{\mu} = T^{\alpha}_{\hphantom{\alpha}\mu\alpha}$$
for the metric teleparallel case of $Q_{\alpha\mu\nu}=0$, and
$$R(\mathop\Gamma\limits^{(0)})+\mathbb Q +  \mathop\bigtriangledown\limits^{(0)}{\vphantom{\bigtriangledown}}_{\alpha}\left(Q^{\alpha}-\tilde Q^{\alpha}\right)=0 \quad \mathrm{where} \quad {\mathbb Q}=\frac14 Q_{\alpha\mu\nu}Q^{\alpha\mu\nu}-\frac12 Q_{\alpha\mu\nu}Q^{\mu\alpha\nu}-\frac14 Q_{\mu}Q^{\mu}+\frac12 Q_{\mu}\tilde Q^{\mu}$$
$$\mathrm{with}\quad   Q_{\alpha}= Q^{\hphantom{\alpha}\mu}_{\alpha\hphantom{\mu}\mu}\quad \mathrm{and} \quad\tilde Q_{\alpha}= Q^{\mu}_{\hphantom{\mu}\mu\alpha}$$
for the symmetric teleparallel one of $T_{\alpha\mu\nu}=0$. Since the Levi-Civita ($\mathop\bigtriangledown\limits^{(0)}$) covariant divergences are nothing but boundary terms, the Lagrangian densities of $\frac12 \mathbb T$ or $\frac12 \mathbb Q$ produce the same equations as the Hilbert one $-\frac12 R$, hence the names of Teleparallel GR-Equivalent (TEGR) and Symmetric Teleparallel GR-Equivalent (STEGR) respectively.

For sure, in every part of physical science there is a certain level of convention at play. One can imagine the very same nature being described in an absolutely different language. People have managed to practically build the Euclidean geometry and the notion of  Newtonian time, and then still with some extra conventions while introducing forces and masses, we've got the classical mechanics. If there existed a civilisation on a neutron star, they might have formulated their mechanics in terms of anisotropic geometry, or with a repelling force from the physical axis, or in an even more different way. Obviously, it would be a type of Aristotelian physics, with their "Earth" in the centre of the world, but for them this picture might have been the most convenient one for a long time, even with advanced mathematical models of physical motion.

However, the situation with teleparallel equivalents is different. In reality, we always need the metric, and therefore the usual curvature is always there, even if we do not mention it explicitly. Then, for reproducing the very same theory prediction-wise, and contrary to the Occam's razor, we introduce some new geometric structures, absolutely unobservable and esoteric. I would say, it doesn't make sense unless we are to use the new formulations for novel modifications of gravity in which the new entities are to become physical. However, the modifications are almost always rather problematic \cite{meprobl}. A recent Newer GR case \cite{unimT} is basically nothing but a reformulation of unimodular gravity, and therefore it does not serve this kind of purpose.

Philosophers might talk about underdetermination in this respect \cite{philosophy}. One might object this attitude by saying \cite{Laur} that a similar situation happens in the modern formulation of electrodynamics. However, it rather relates to GR itself than to its teleparallel reformulations. Indeed, if we do not insist on quantisation or on a nice action principle, whatever gauge we choose or don't choose at all does not make any difference. The physical quantity is the electromagnetic field strength. The gauge freedom of choosing the vector potential is immediately understood as redundancy of our description. Moreover, representation of the field strength in terms of the vector potential, $F_{\mu\nu}=\partial_{\mu}A_{\nu} - \partial_{\nu}A_{\mu}$, can be taken as just a convenient solution for the "homogeneous" part of the Maxwell equations. The same is in GR: the physical object is the metric tensor, while its components are only for our convenience of working with numbers.

The teleparallel equivalents are different, as long as their proponents insist on some physical reality behind the flat parallel transport and its torsion and/or nonmetricity tensors. If we treat the teleparallel structures as nothing but an abstract tool which might or might not be convenient, then I can agree with the conventionalism opinion of the paper \cite{Laur}, even though it looks a bit strange in this case, to simultaneously insist \cite{full} on the fully covariant approach to such purely technical instruments. And this is definitely not the case if one claims that the teleparallel gravity has allegedly been able to separate gravity and inertia \cite{textbook}, by simply pronouncing the zero-spin-connection or "proper" tetrads the inertial frames. One could do the same in electromagnetism: instead of a gauge-breaking mass term which would make the full information about the vector field physical, one could add a total-divergence gauge-breaking term to the Lagrangian density, like $\partial_{\mu} A^{\mu}$ or $\square A^2$, thus not changing anything predictable but getting an opportunity of insisting on having something physical in the gradient part of the vector potential.

In other words, a teleparallel inertial frame on the spacetime manifold appears as just an arbitrary choice of a family of new autoparallel curves which show no deviation. Those are orthonormal frames in TEGR and coordinate ones in STEGR. However, in unmodified models, they bear no influence on the predictable outcomes. Actually, we can do the same without any teleparallel philosophy at hand: let's simply say that the physical world is a Minkowski spacetime with the dynamical metric being just a tensor field on that, and then we have the notion of inertiality with respect to Minkowski. Pragmatically, it is just the same as STEGR, with the full arbitrariness in how to relate the real world to the imagined Minkowski space and its emergent Poincar{\'e} symmetry, i.e. with the freedom of choosing the coordinates which are assumed to bring the Minkowski metric into the standard form of $\eta_{\mu\nu}$.

Note that, in a topologically trivial space, one can choose any global coordinates and call them the Cartesian ones thus building the coincident gauge of STEGR on it. Of course, the global coordinates I have in mind are those in the full mathematical meaning, therefore prohibiting all the ugly constructions like spherical coordinates, for example. Recall also that one of the early objections to GR was that the requirement of general covariance does not make any sense since every model can be rewritten in a way respecting it \cite{Kretsch}. The catch is that GR is a geometric theory of a pseudo-Riemannian manifold with no additional structures on top, and this is a non-tirival statement, unlike the general covariance per se. Going for symmetric teleparallel frameworks removes this aspect and makes the Kretschmann's objection \cite{Kretsch} fully valid. Whatever strange partial derivative expression in terms of the metric components we see, one can always say that it is a tensorial thing written in some special coordinates of the "coincident gauge".

All in all, every metric-affine framework introduces a novel geometric structure on the spacetime manifold. It can be represented as an independent connection, or as a tensor describing the difference of this connection from the Levi-Civita one. The latter option does not only give us a better intuition of having something new, but very often it is also very convenient for calculations, see for example a recent paper \cite{metraff}. In the special case of teleparallel theories, the novel geometric structure can also be given in terms of preferred tangent space bases, those corresponding to zero connection coefficients. It looks a bit similar to a would-be preferred vector potential in electrodynamics. However, this very approach makes it very unnatural to insist on the equations of motion being invariant under changes of a basis, or under gradient shifts of the vector potential. It means then trying to make physics out of something which is simply not in the equations at all. Yet, this is precisely the fine-tuning of the teleparallel equivalents of GR, with most generalisations being problematic.

\section{On the dreams of conserved energy and quantisation}

One of the teleparallel gravity benefits people often claim to have \cite{enercan, textbook} is an alledged solution for the energy problem. To start with, it is not even clear what was the problem. In GR, there is simply no covariant notion of conserved energy, nor any other usual conservation laws related with Minkowski space symmetries, and it shouldn't be there for the lack of such symmetry. Full stop. Of course, conserved quantities do exist in speical cases when appropriate Killing vectors are available, or as global charges if the spacetime asymptotically possesses such symmetry, and one can even go for quasi-local constructions \cite{quasloc}. It does not mean though that we can expect that, in general situations, meaningful conservation laws would suddenly appear out of nowhere. Actually, the same is true \cite{glcha} of non-Abelian Yang-Mills theories, too.

The "problem" is more about not satisfying our personal habit of having conserved energy and momentum, and it does not have anything to do with real physics. In a sense, when people invest lots of effort \cite{enercan, enertel} into looking for a "correct" definition of some familiar word in situations when there is no reason for it to be applicable, it is a very strange picture to see \cite{pamphlet}. A definition cannot be correct or incorrect, it is our free choice. And if there was some independent objective meaning behind, there won't be any task of looking for it. If we take ${\mathcal T}^{\mu\nu}$ as a description of conservation following from spacetime symmetries, there is no such symmetry in case of gravity. And even when the symmetry is there, the definition is not unique \cite{bello}. There is nothing canonical in any "canonical" \cite{enertel} expression, except for desirable options such as gauge invariance when needed, or absence of unnecessary complications, or convenience of defining the angular momentum \cite{bello} indeed. Any particular result of deriving a conserved tensor rather reflects a way of proving the Noether theorem than any objective canonicity.

There is one particular (symmetric) form of the energy-momentum tensor of matter fields which enjoys another physical meaning. It is the source for gravitational interactions. But then it does not have to yield conserved charges, outside the very local, i.e. almost Minkowskian considerations. Moreover, it cannot do that, whatever generalisation in the realm of GR we try, unless we go for the beautiful idea of Levi-Civita to call minus the Einstein tensor the energy-momentum tensor of gravity, thus having a conserved zero in total. This is just because any expression of the ${\mathop\bigtriangledown\limits^{(0)}}_{\nu} {\mathcal T}^{\mu\nu}=0$ shape for a non-zero tensor does not in general lead to an integral conservation, let alone the absence of symmetry to justify it, in the first place. On the other hand, the lower order of the differential equation still perfectly serves its role of helping us to understand and solve equations, think for example about the energy density behaviour $\dot\rho + 3H (\rho + p)=0$ in cosmology.

In other words, there is not enough structure in the objective geometry to provide us with the usual conservation laws. What the teleparallel equivalents do is constructing some extra things, in nonpredictable and unobservable ways. In my opinion, it is no better than the non-covariance of numerous classical options. If to treat ${\mathcal T}^{a\mu}=h^a_{\nu}{\mathcal T}^{\nu\mu}$ as four vectors, then their vanishing divergences, if found so, do have an integral conservation feature. Moreover, since the teleparallel approach does have a tetrad $e^a_{\mu}$ with vanishing connection coefficients in its basis, then using this unobservable preferred basis as an "inertial frame", a zero partial-derivative divergence of $\sqrt{-g}\cdot e^a_{\nu}{\mathcal T}^{\nu\mu}$ can be given a perfectly covariant interpretation \cite{textbook}. And in case of symmetric teleparallel gravity, these conservation properties can be even viewed as representing the very familiar symmetry of the fictional Minkowski space (of $\zeta^a$ Cartesian coordinates) coming into play. On the other hand, whatever non-covariant law we have, it can be pronounced covariant by simply saying that every partial derivative was a covariant one in the coincident gauge. 

In terms of the most recent teleparallel language, one can reproduce the Einsteinian energy-momentum complex by defining the "canonical frames" \cite{canon} as those in which what they call the metric energy-momentum tensor vanishes \cite{parcosm, enercan, canon}. Their claim is that this condition is a coordinate-independent definition of the frames. It is however, to a large extent, a bit of cheating. The tensorial nature is only due to formal rewriting of an intrinsically non-covariant thing in covariant terms which can always be done \cite{Kretsch}. In symmetric teleparallel, our working coordinates are free to be chosen, but the former non-covariance is now given in terms of the flat parallel transport with preferred (i.e. Cartesian) coordinates $\zeta^a$ which, in the case of STEGR, are not predictable either, and therefore not verifiable experimentally. 

As long as we stay with the GR-equivalent versions of gravity theories and the usual minimal coupling of matter to the metric, there is nothing observable about the teleparallel structure. Just to be clear, I do not canonise the minimal coupling either. In my view, unlike for some other people, even in case of gravity with torsion, keeping the partial derivatives in the definition of the electromagnetic field strength tensor in terms of the vector potential is considered as a part of the minimal coupling.

Teleparallel gravity then makes all the same predictions as GR itself, though with a more complicated geometric structure in the foundations, therefore going against the Occam's razor. As a benefit, they put forward a covariantly conserved energy, with no good argument for why it should exist, to start with (recall the lack of corresponding symmetry in general relativistic spacetimes). Their go for defining {\bf the energy} at any price is somewhat akin to Occam's wish to accept existence of god without proving, despite his own razor. It might be an explanation of love for the word "canonical" observed in some proponents of the modern developments in teleparallel gravity.

Having said all that, I must admit that there is an interesting motivation to look for the energy. Namely, it can be considered as one of the most important observables in quantum mechanics. One has to bear in mind though that it is nothing but common jargon of quantum physicists who tend to call an arbitrary self-adjoint operator an observable, while the Hamiltonian operator is also important because it is the thing which defines the Schr{\"o}dinger evolution. The point is that, if it was anything really observable, there won't be any problem of defining it, one way or another. And taking the canonical gravity approach, this "very important observable" must just be zero, with no issue of defining it indeed. It actually touches upon just one particular aspect of a much deeper issue. I would say, the gravitational and the quantum are absolutely incompatible with each other, already in the way how they treat the notion of time.

This is a problem, for sure, as long as we want to have a full self-consistent theory of the physical world. The common conclusion that one has to modify gravity for that is hard to accept though. Why not change the quantum mechanics? We often tend to forget that quantum physics is very problematic in itself, up to just making no good sense when taken as a fundamental theory. It works perfectly as an effective model for predicting probabilities of various outcomes in laboratory experiments. However, one cannot say the same about our understanding of what is going on in reality. I mean the infamous measurement problem and the quest for a proper interpretation of quantum mechanics. It is a pity that most textbooks simply ignore the problematic nature of quantum mechanics, with a nice exception being in the wonderful book of S. Weinberg \cite{QM}.

The standard Copenhagen interpretation involves a classical observer with a measuring device, and they somehow do not belong to the quantum world. It is all right for an effective theory which does not pretend to be a full description of the world but rather works with observations in a setup of limited information available. Then the Newtonian absolute time and big objects beyond understanding, and maybe even the paradoxes of Schr{\"o}dinger cats and effective collapse of wave functions, are something natural and to be expected. However, can we also apply it to gravity, let alone cosmology, with no objects outside of the supposedly quantum world? Probably, not. In my opinion, the only conservative attitude is the Many Worlds Interpretation, for this is an inevitable outcome in case the quantum laws apply to everything in the universe. Of course, there are issues with it, too, like the task of really deriving the Born rule, and also the spirit of Occam being summoned again, to say the least.

I must also mention that there are different opinions on the very basic question of what is quantum, as opposed to classical. One can take a $\psi$-epistemic stance and say that we simply do not have any proper theory of the quantum world, but a mere description of our knowledge of it. Then it is possible to say that almost everything which is usually considered quantum is just the same indeterminacy we might have in the classical physics too, except the incompatible measurements \cite{quantclass}. It is not fully honest because in classical physics we do assume that in principle there does exist the full dynamical information, even if not accessible for us. On the other hand, taking the Many Worlds Interpretation in the $\psi$-ontic manner, I may claim that there is nothing really quantum at all. There are no particles or fields in reality, but just the "quantum states" evolving purely deterministically, and all the incompatible measurements are then related to our attempts of measuring non-existent things.

Another option people often discuss is about objective collapse models which go beyond quantum mechanics by adding non-linear terms to the equations. It is a worrisome game to play since it is only due to linearity of quantum mechanics that the processes which do look like an instantaneous collapse of the wave function do not lead to observable and usable violations of relativistic causality. They often claim that the nonlinear terms are necessary for cosmology because the Many Worlds Interpretation is problematic for predicting the cosmological perturbations \cite{qorig, qorig2} since, in absence of any outer world, there is no mechanism for the decoherence to create an inhomogeneous observed universe from the exactly homogeneous initial quantum state of inflation. This is however wrong. Unlike tunnelling from nothing, inflation has never assumed an ideally symmetric initial state. It is rather used as a natural mechanism for evolving towards a state which is symmetric to a high precision. This is not to say that I think that, to the contrary, a superposition of all possible classical worlds with various distributions of matter is free of any problem. However, I don't really see any better option for now.

Leaving the foundational issues of quantum mechanics per se behind, as I've already mentioned above, it has a complicated relationship with relativistic causality. In a relativistic field theory, relying on the spacelike-separated free-field commutators vanishing due to a particle and an antiparticle in opposite directions cancelling each other, it does not provoke much more than purely epistemic thinking towards the foundations, so that it is not surprising at all that finally we end up with an $S$-matrix only. Even ignoring these issues, there is no mathematically rigorous definition of any realistic Quantum Field Theory \cite{Ded}. Despite the beautiful intuitive picture of Wilsonian renormalisation, we still severely lack proper mathematics, and maybe also parts of physics. With all that, I would not worry about technical issues such as non-renormalisability of quantum GR, for there is no reason to expect that the Nature must be nice to our abilities.

In absence of available experiments in quantum gravity, what is nice about the standard Quantum Field Theories, and even about String Theory to some extent, is that they are very conservative, and therefore do not leave us with too much freedom. On the other hand, they might be just wrong at the level we need for building the correct quantum gravity, if any. Trying the Loop Quantum Gravity (LQG) approaches seems very reasonable to me, for they at least care about mathematical rigour and background independence. At the same time, when taking quantisation rigorously, there is a lot of ambiguity in how to do a corresponding deformation. In particular, with the LQG techniques, they can do it with no anomaly at all, even for the bosonic string in arbitrary dimension \cite{Thomas}. The price to pay is total abandonment of any particle-excitation picture, up to observables and semiclassical limits not being quite clear. However, it might indeed be a way to go.

This distraction has been a bit too long. All I wanted to say is that the needs of the standard quantum physics might not be a proper motivation for trying to construct something in GR which is not natural for it. What the teleparallel-equivalent reformulations do is nothing but an arbitrary choice of some preferred bases with respect to which the otherwise unnatural things become well-defined, kind of reviving the notion of inertial frames and conserved quantities. However, at least for the symmetric teleparallel incarnation, it does not even need any teleparallel philosophy. One might simply say that the metric is just a dynamical field on a Minkowski spacetime. 

In particular, the recent paper \cite{quantgrav} can be seen as a fundamentally STEGR-type quantisation of gravity. There is no background independence in their construction whatsoever, even though the authors might object this claim of mine. They take the one and only vacuum, the Minkowski spacetime, as a sacred entity for quantum theory, so that all the usual things such as conserved energy and even $S$-matrix seem to be well-defined, and then all other possible spacetimes are treated simply as coherent states. The problem is the same as for the genuinely teleparallel approaches. This Minkowski is an abstract and unobservable construction. Even for a spacetime of trivial topology, there is an infinite freedom of choosing those global coordinates which correspond to the Cartesian ones on the imagined Minkowski structure.

\section{Conclusions}

The teleparallel gravity models are being very actively discussed nowadays, in particluar with the ideas of defining conserved energy and inertial frames in the GR-equivalent ones. Fully understanding the possible motivations from the quantum physics side, I still insist that these notions are absolutely unnatural for the classical gravitational physics. One has to bear in mind that the claims that energy is the most important observable are related to the unfortunate habit of calling self-adjoint operators observables, and to the Hamiltonian picture which in itself goes against the general relativistic view, for there is no objective notion of time there.

The notion of objectively preferred ("inertial", "proper", "canonical", you name it) frames contradicts the equivalence principle, while the theory of general relativity is fully compatible with it, and no experimental detection of its violation is around. Therefore it goes rather as introducing something unverifiable, unless we modify the dynamics of gravity. At the same time, if to think on the practical side, it sounds very strange when people say that something is observable and is very important, at the same time having troubles to even define what it is, in a theory which works perfectly. The point is that the energy is nothing directly given in observations. It is an artificial construction, though a very important one indeed, which uses symmetry of our old theories with respect to time translations. Once the observable geometry has lost such symmetry, there is no conserved energy either.

The Teleparallel Equivalents of GR do introduce a new geometric entity on top of the usual (and I insist, observable) Riemannian geometry. Equivalence to GR requires a fine-tuning of the action functional such that the choice of the flat parallel transport is absolutely arbitrary, in that it does not influence the equations of motion at all. It is very interesting for doing new modifications or quantisations of gravity, or for other foundational studies. However, as a solution to the "problem of energy" it is not much more than an arbitrary choice of what to call inertial frames for defining the usual "observables" or "charges" with respect to them, with no proper justification and with no window for physically detecting this flat structure, not even in principle.

Coming back to the question in the title, in my opinion, there is no geometric trinity of gravity. If gravity is not considerably modified, its only really physical representation is in terms of the observable and predictable pseudo-Riemannian geometry. The teleparallel approaches to GR do introduce some extra flat structures on top of it. Those are neither observable nor predictable. Formally, they allow us to use the flatness of the new connection for constructing familiar notions and for translating the gravitational equations into the language of this flat geometry and its differences from the real one. However, it then entails mixing the physical observations with arbitrary creations of our imagination. I would not call it physics.

{\bf Acknowledgements}. I am grateful to the organisers of the 11th Mathematical Physics Meeting, September 2024 in Belgrade, Serbia. It was a very interesting conference with many helpful discussions. 

I would also like to thank Tomi Koivisto for private email correspondence. It was his reaction to my paper which I had written for proceedings of the Metric-Affine Frameworks for Gravity conference in Tartu 2022 \cite{meW}, and he brought the idea of energy being the most important observable to my attention. Needless to say, his opinion is pretty much different from what is presented above.

\end{document}